# 25. Y Dwarfs: The Challenge of Discovering the Coldest Substellar Population in the Solar Neighborhood

Sandy K. Leggett[1]

(1) Gemini Observatory, Northern Operations Center, Hilo, HI, USA

**Sandy K. Leggett**
**Email:** sleggett@gemini.edu

## Abstract

Stars form in the Galaxy with a wide range in mass. If the mass is below 7% of the Sun's, then the object does not become hot enough for stable hydrogen burning. These substellar objects are called brown dwarfs. Maps of the sky at infrared wavelengths have found large numbers of brown dwarfs. However only 24 objects have been found (as of April 2017) that are cold enough to be classified as "Y dwarfs": these have atmospheres that are cooler than 500 K (or ~200°C, 400°F) and have masses only 5–20 times that of Jupiter. The coolest Y dwarf currently known, discovered in 2014, has a temperature around freezing, has a mass of about 5 Jupiter masses, and is only 2 pc away from the Sun. These small and cold objects are faint and difficult to find. This chapter describes the discovery and characterization of the Y dwarfs. Finding more of these very cold planet-like brown dwarfs will require an as-yet unplanned space mission mapping large areas of sky at wavelengths around 5 μm.

## Introduction

The stars that form in galaxies have a wide range in mass (Salpeter *1955*; Miller and Scalo *1979*; Kroupa *2001*). Stars can form with a mass as large as 150 times the mass of our Sun (Zinnecker and Yorke *2007*), while the lower mass limit for formation of a starlike object is currently not known. Chabrier (*2003*) notes that turbulence in protostellar clouds appears to generate structures with a mass smaller than that of Jupiter. Observations and theory suggest that very low-mass objects can form either like a star from molecular cloud fragmentation (Bowler and Hillenbrand *2015*) or like a planet via stellar disk instabilities (Li et al. *2015*). Disk-instability objects may initially have higher temperatures than molecular cloud clumps, and models have been calculated for so-called "hot-" and "cold-start" formation (Spiegel and Burrows *2012*). Traces of the start condition disappear after about 100 Myr, however making it extremely difficult to distinguish between a cold very low-mass starlike object and an isolated gas-giant planet.

Independent of the formation question, it has been recognized since the 1960s that there is a mass below which the core of the object does not become hot enough for stable hydrogen burning (Hayashi and Nakano *1963*; Kumar *1967*). These objects, known as brown dwarfs, have a mass less than 7% of our Sun's and a radius similar to that of Jupiter (Baraffe et al. *2002*; Saumon and Marley *2008*). Because there is no sustained source of energy in their interiors, brown dwarfs cool with time. During the first Gyr of a brown dwarf's life, the luminosity decreases by a factor of ∼100, and 1–75 Jupiter-mass brown dwarfs cool to effective temperatures ($T_{\text{eff}}$) of ∼200–2000 K, respectively. Small and cold objects like brown dwarfs emit very little light and are difficult to find unless they are nearby. This chapter describes the discovery and characterization of the coldest examples known at the time of writing of the brown dwarf or substellar group. They are called Y dwarfs.

## Discovery of the Y Dwarfs

Stars and brown dwarfs are not blackbodies, but the laws of blackbody radiation can be used to estimate their spectral energy distribution. The Stefan-Boltzmann law states that the power emitted is proportional to $T^4$, where $T$ is the blackbody temperature in Kelvin. Similarly, as a brown dwarf cools, its brightness decreases. Wien's law for blackbody radiation states that the wavelength $\lambda$ at which the blackbody emission peaks is $\lambda = 0.0029/T$ m. Therefore a 1000 K blackbody is brightest at 3 μm and a 300 K blackbody is brightest at 10 μm. For cool brown dwarfs, most of the light emerges in the infrared region of the electromagnetic spectrum.

Significant numbers of brown dwarfs started to be found in 1999, in infrared surveys of the sky. First the so-called L- and T-type brown dwarfs, with $T_{\text{eff}}$ ∼ 2000 and 1000 K, respectively, were found by near-infrared surveys (Burgasser et al. *1999*; Kirkpatrick et al. *1999*; Strauss et al. *1999*; Warren et al. *2007*), and then very late-type T and Y dwarfs with $T_{\text{eff}}$ ∼ 500 K were found by mid-infrared sky surveys (Cushing et al. *2011*; Mainzer et al. *2011*). Figure 1 is a color-magnitude diagram which shows imaging data for L, T, and Y dwarfs. The intrinsic flux emitted at 4.5 μm is plotted on the $y$ axis of the left panel, and the model-implied $T_{\text{eff}}$ K is shown on the $y$ axis of the right panel. As $T_{\text{eff}}$ decreases, the absolute brightness of the brown dwarf decreases, and more light emerges at 4.5 μm compared to the shorter wavelength $J$ (1.25 μm) and 3.6 μm bands.

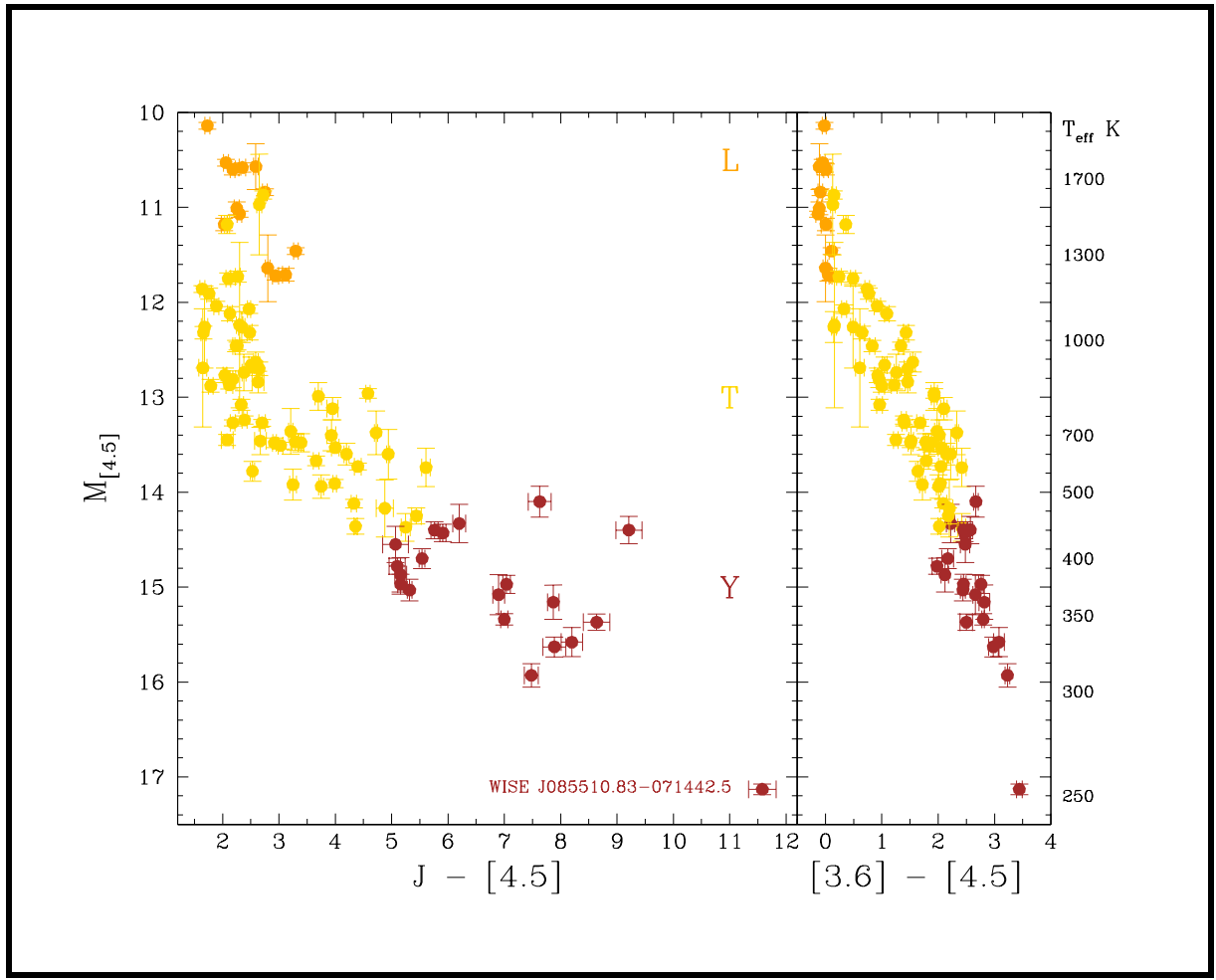

**Fig. 1** The *y* axis gives the absolute [4.5] magnitude measured for a sample of L (orange), T (yellow), and Y (brown) dwarfs. The left panel shows these data as a function of *J* (1.25 μm) − [4.5] color, and the right panel as a function of [3.6] − [4.5] color. Brown dwarfs that are unresolved multiple systems are brighter for a given color (The data are from Leggett et al. *2002*, *2017*, and references therein).

The observed spectrum of a T or Y dwarf is shaped by broad bands where molecules in the brown dwarf's atmosphere absorb light. The strongest features are due to $H_2$, $H_2O$, $CH_4$, and $NH_3$ absorption (Morley et al. *2014*). The transmissivity of Earth's atmosphere is poor where $H_2O$, $CH_4$, $CO$, $CO_2$, $N_2O$, $O_2$, and $O_3$ absorb light (Lord *1992*). Conveniently, much of the light that emerges from a brown dwarf atmosphere between its absorption bands can be detected by ground-based observatories because those wavelengths coincide with clear regions of the Earth's atmosphere. Figure *2* shows modeled spectra of brown dwarfs with the transmission spectrum of the Earth's atmosphere overlaid.

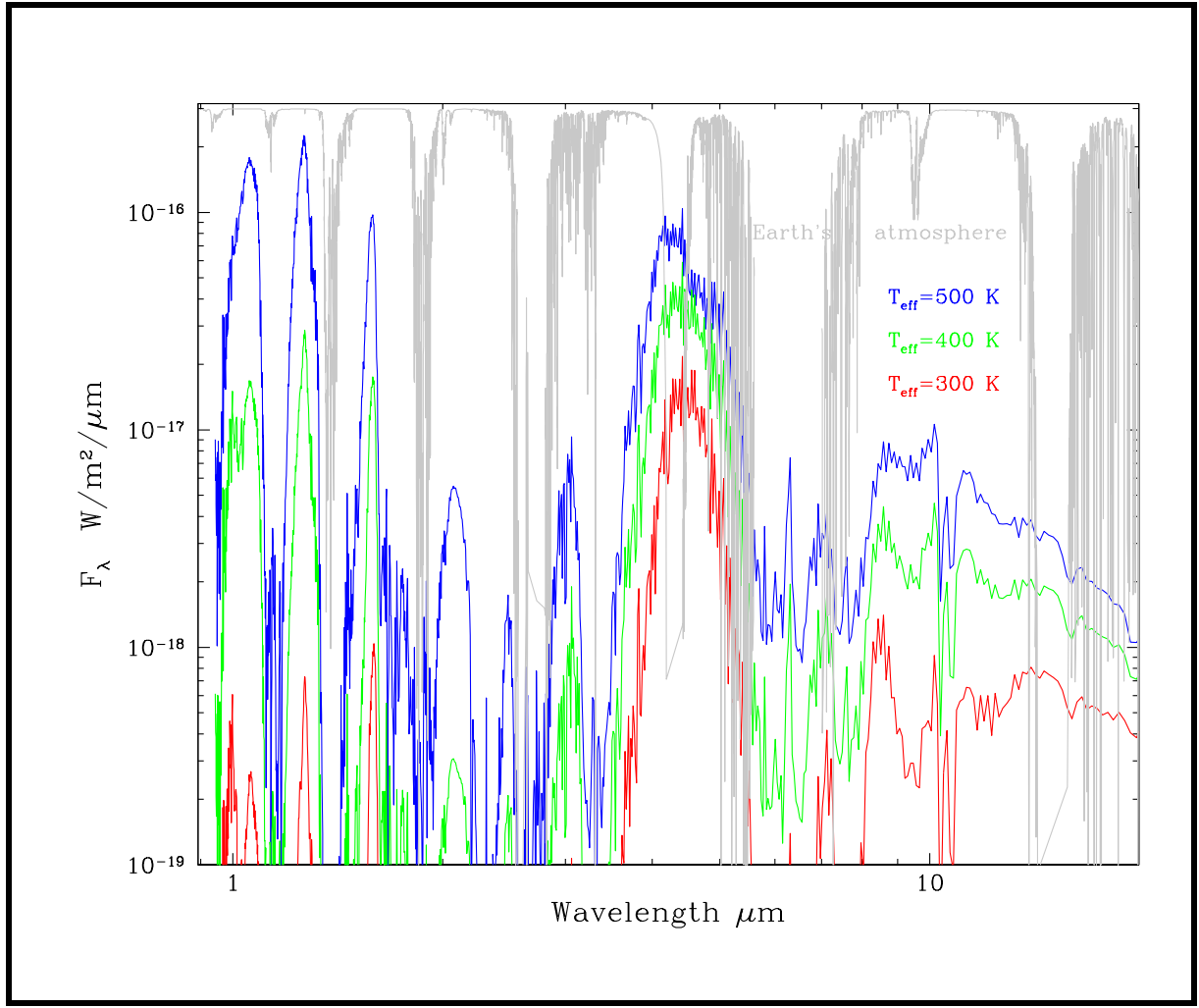

**Fig. 2** Synthetic spectra for brown dwarfs with $T_{\text{eff}}$ = 500, 400, and 300 K are shown as blue, green, and red lines, respectively. The spectra are generated by Tremblin et al. (*2015*) nonequilibrium chemistry models for a cloud-free atmosphere with solar metallicity and a surface gravity $g = 10^{4.5}$ cm s$^{-2}$. The gray line is the relative transmission of the Earth's atmosphere from Lord (*1992*). Note that the axes are logarithmic.

At wavelengths longer than ∼3 μm, thermal emission from the atmosphere and telescope are significant, and ground-based observations become difficult. The transparency of the Earth's atmosphere is also poor at 4.5 μm, where approximately half the light from a cold brown dwarf emerges (Fig. 2). For these reasons the coldest brown dwarfs known, those with $T_{\text{eff}}$ < 500 K, were detected by space missions, at $\lambda \approx$ 4.5 μm. The cameras used were the Infrared Array Camera (IRAC; Fazio et al. *2004*) on board the *Spitzer* telescope and the imager on board the *Wide-field Infrared Survey Explorer* (*WISE*; Wright et al. *2010*). Despite the high sensitivity of *Spitzer* and *WISE*, the cold brown dwarfs are so faint that only those closer than ∼20 pc can be detected. Twenty-four brown dwarfs with $T_{\text{eff}} \leq$ 450 K, classified as Y dwarfs, are known as of April 2017. A list of these objects, with photometric data and distances where measured, is given in Table 8 of Leggett et al. (*2017*) at http://iopscience.iop.org/article/10.3847/1538-4357/aa6fb5/meta#apjaa6fb5t8.

# Why "Y"? Spectral Classification

Most stars can be classified by the one of the letters in the OBAFGKM sequence (Cannon and Pickering *1912*; Morgan and Keenan *1973*). Additional letters are used for other types of stars, galaxies, and nebulae. Kirkpatrick et al. (*1999*) reviewed the available letters that could be used for a new spectral class and found that only H, L, T, and Y were unambiguous. The new objects that were found in infrared sky surveys were cooler than M-type stars, and so the letters L, then T, and then Y were used as cooler and cooler objects were discovered. Each letter class is subdivided into numbers so that the boundary between T and Y, for example, is represented by T8, T9, Y0, and Y1.

The L dwarfs are spectrally similar to M dwarfs. The near-infrared spectra of T dwarfs are markedly different from the L dwarfs because they show absorption by $CH_4$. The Y class is spectrally similar to T dwarfs, but the width of the flux peak at $\lambda \approx 1.25$ μm (see Fig. 2) is narrower, implying stronger absorption by molecular bands at shorter and longer wavelengths. Cushing et al. (*2011*) and Kirkpatrick et al. (*2012*) have used the width of this flux peak to define a sequence for the end of the T and start of the Y class. Very few Y dwarfs are known however, and it is possible that the spectral classification of the brown dwarfs with $T_{eff}$ < 500 K will need to be redefined in the future, when more examples are known (although finding more is difficult as described in the last section of this chapter).

# The Structure and Atmosphere of Cold Brown Dwarfs

The interiors of brown dwarfs are composed of ionized hydrogen and helium, and this structure is supported against gravitational collapse by electron degeneracy pressure and Coulomb forces between the ions (Burrows et al. *2001*). After ∼200 Myr a brown dwarf has contracted to a size similar to that of Jupiter and does not become significantly smaller (Burrows et al. *1997*). The interior of a brown dwarf is fully convective, and cooling proceeds adiabatically (Saumon and Marley *2008*). The thermal energy is emitted through the radiative atmosphere. The cooling of the brown dwarf is controlled by the energy loss through the molecule-rich atmosphere.

Three-dimensional simulations of convection in rotating brown dwarf and giant planet interiors show that significant circulation is generated (Showman and Kaspi *2013*). Horizontal temperature variations can result in flux variations of a few percent with periodicity similar to the rotation period of the brown dwarf. Stratified turbulence can generate vortices and storms, and the circulation can support the formation of patchy clouds. It is likely that both thermal and cloud cover fluctuations are important sources of variability in giant planets and brown dwarfs. Variability of a few percent has been detected in Y dwarfs at mid-infrared and possibly near-infrared wavelengths (Cushing et al. *2016*; Leggett et al. *2016*).

Various species condense in these cold atmospheres, forming cloud decks. For T dwarfs with 500 < $T_{eff}$ K < 1300, the condensates consist of chlorides and sulfides (e.g., Tsuji et al. *1996*; Ackerman and Marley *2001*; Morley et al. *2012*). As $T_{eff}$ decreases further, the next species to condense are calculated to be $H_2O$ for $T_{eff} \approx 350$ K and $NH_3$ for $T_{eff} \approx 200$ K (Burrows et al. *2003*; Morley et al. *2014*). Observations support the presence of chloride and sulfide clouds in T dwarf atmospheres and possibly even in deep layers of Y dwarf atmospheres (Leggett et al. *2013*, *2016*). The presence or absence of water clouds in the coolest known Y dwarfs is an active area of current research.

# Physical Properties of Cold Brown Dwarfs

The surface gravity $g$ of a star or brown dwarf is defined as $g = GM/R^2$ where $G$ is the gravitational constant, $M$ is the mass, and $R$ is the radius of the star or brown dwarf. Evolutionary models show that $g$ constrains mass, because the radii of brown dwarfs do not change significantly after about 200 Myr and are within 25% of a Jupiter radius (Burrows et al. *1997*). Also, the cooling curves as a function of mass are well understood, so that $T_{eff}$ combined with $g$ constrains both mass and age (Saumon and Marley *2008*). Evolutionary sequences are shown in Fig. 3.

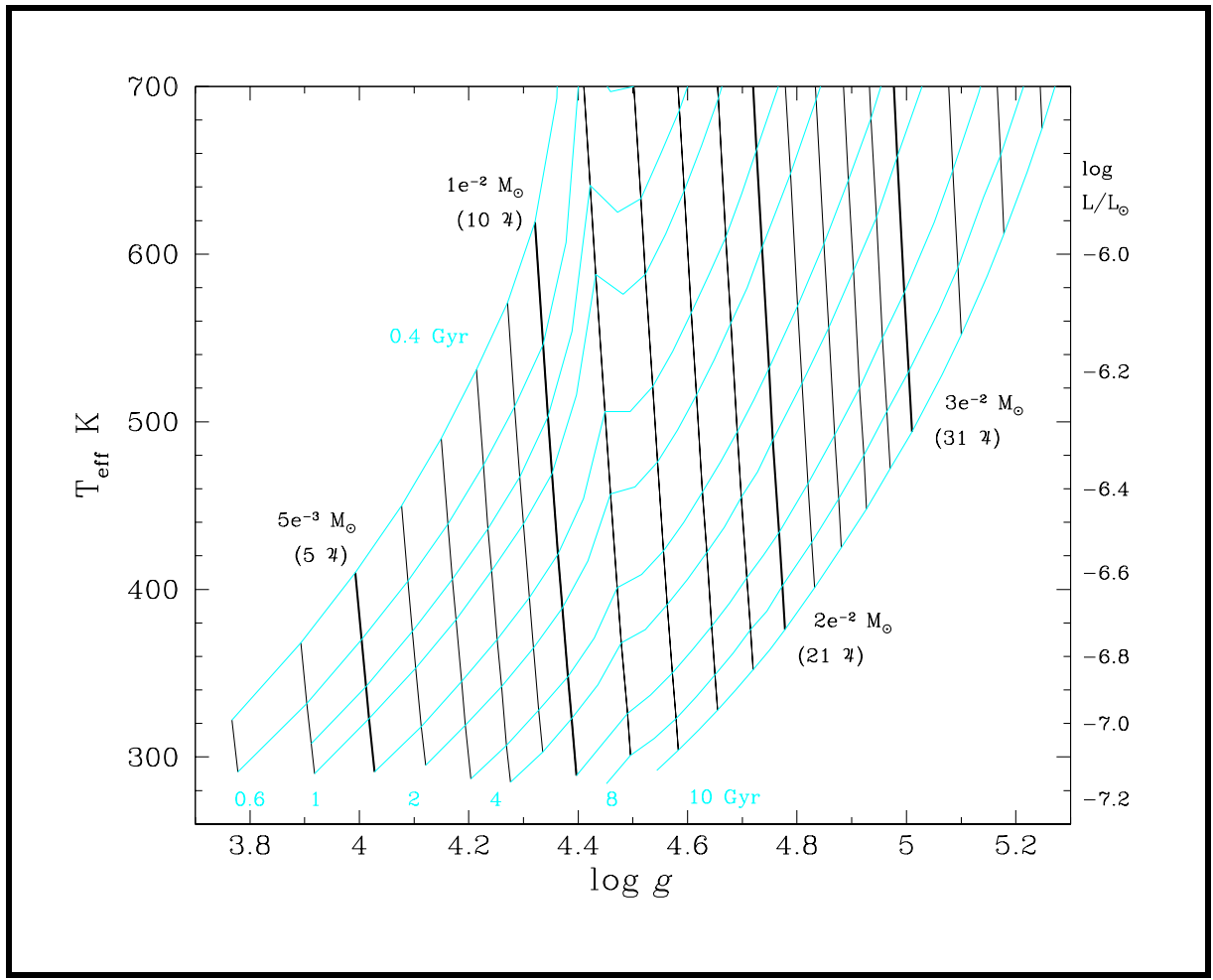

**Fig. 3** Evolutionary sequences for cool brown dwarfs for an age range of 0.4 to 10 Gyr from Saumon and Marley (*2008*). Black lines are constant mass, cyan lines are constant age. Mass and gravity $g$ are correlated as the radii of these brown dwarfs are all ≈1 Jupiter radius. Legends indicate mass and age. One Solar mass = 1047 Jupiter masses.

The correlations between $g$ and mass, and $T_{eff}$ and age, mean that even for the predominantly isolated brown dwarfs in the solar neighborhood, mass and age can be estimated if $T_{eff}$ and $g$ can be determined, for example, by fitting the spectral energy distribution calculated by atmospheric models to observations. Models of brown dwarf atmospheres have advanced greatly in recent years. Opacities have been updated for $CH_4$, $H_2$, and $NH_3$ (Yurchenko et al. *2011*; Saumon et al. *2012*; Yurchenko and Tennyson *2014*). Models which include sedimentation by various species, i.e., clouds, are available (Morley et al. *2012*, *2014*). Models are also available which include a disequilibrium treatment of the atmospheric chemistry, which is a result of vertical gas transport (Saumon et al. *2000*; Zahnle and Marley *2014*). These nonequilibrium chemistry models (Tremblin et al. *2015*) fit the data better than equilibrium models (Leggett et al. *2017*). The models are accurate enough that the physical parameters of the brown dwarf atmospheres can be constrained by comparing observations to synthetic colors and spectra. Leggett et al. (*2017*) find that cloud-free models can reproduce the near-infrared spectrum of the warmer Y dwarfs with $425 \leq T_{eff}$ K $\leq 450$ well, but there is a systematic discrepancy for the cooler Y dwarfs that may be addressed by including $H_2O$ clouds (see Fig. 4).

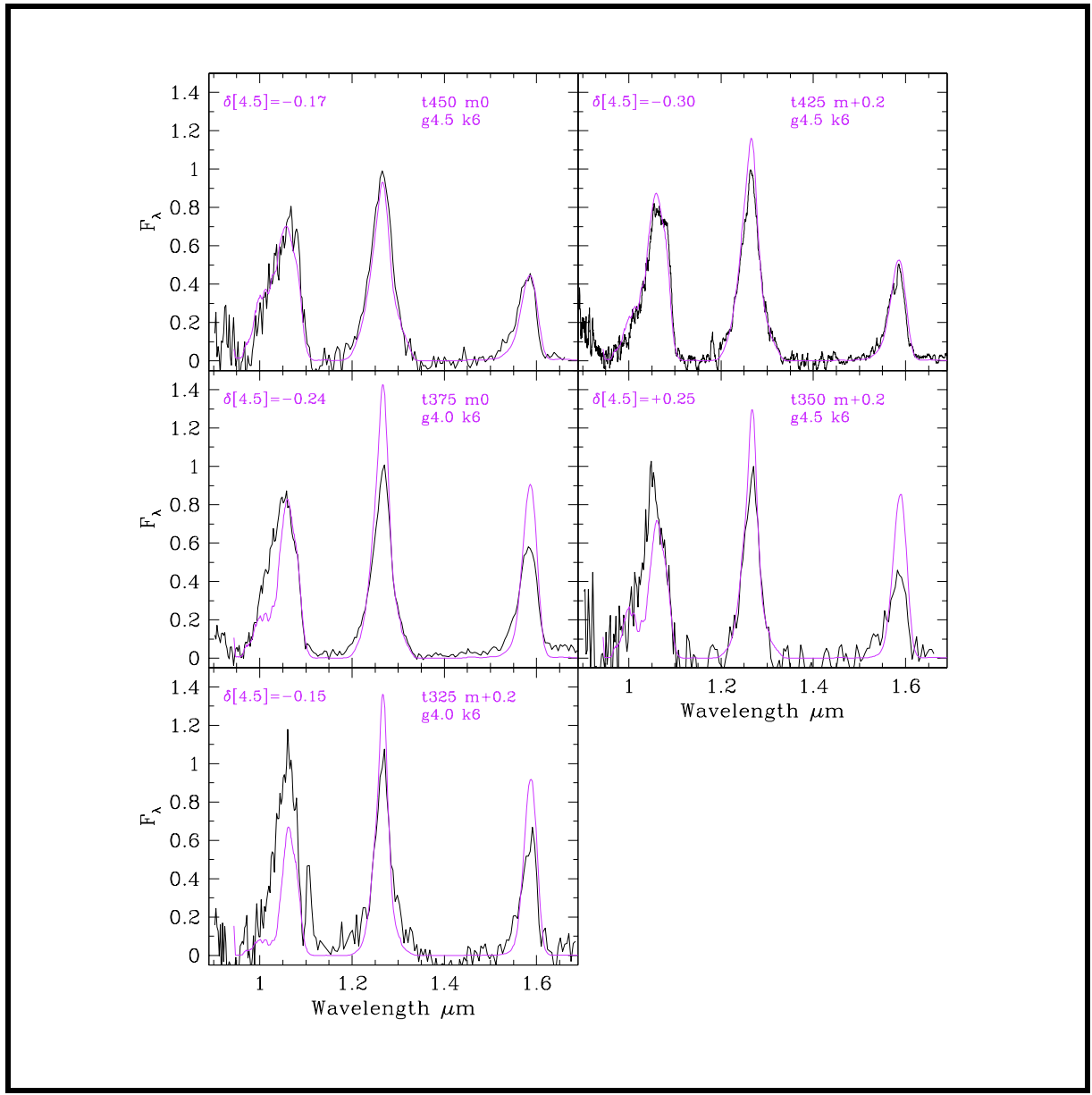

**Fig. 4** Tremblin et al. (*2015*) cloud-free model spectra fits (magenta lines) to observed near-infrared spectra (black lines) for Y dwarfs with a range of $T_{eff}$, from Leggett et al. (*2017*). The legend gives the model $T_{eff}$ as "t," the metallicity [m/H] dex as "m," log $g$ as "g," and log of the diffusion coefficient (for vertical gas mixing) as "k." Also given is the difference between the calculated and observed value of the IRAC [4.5] magnitude, "$\delta$." The fit is good for the warmer Y dwarfs, but for the cooler Y dwarfs, there is a systematic offset such that the 1.05 μm flux is too low and/or the 1.25 and 1.6 μm flux is too high. The discrepancy may be due to the onset of $H_2O$ clouds in the brown dwarf atmospheres, which are not yet included in these models.

Estimates of mass and age for 22 of the 24 known Y dwarfs are given in Table 9 of Leggett et al. (2017) at http://iopscience.iop.org/article/10.3847/1538-4357/aa6fb5/meta#apjaa6fb5t9. Most of these Y dwarfs are ∼15 Jupiter-mass objects with ages and metallicities similar to our Sun, as would be expected for a local sample. Five of the Y dwarfs appear to be younger and have lower mass than the others, with ages less than 3 Gyr and mass ∼8 Jupiter mass. Importantly, the *WISE* discovery of isolated low-mass objects near our Sun suggests that brown dwarfs can form like stars, in molecular clouds, down to masses of only a few Jupiters.

## WISE J085510.83−071442.5

The coolest known Y dwarf as of April 2017 is WISE J085510.83−071442.5 (hereafter WISE 0855). WISE 0855 was discovered by Luhman (2014) as a fast-moving object in multi-epoch *WISE* images. This Y dwarf appears to move quickly against the background stars because it is only 2 pc away. Figure 1 shows that it is one to two magnitudes (a factor of 3 to 6) fainter than the other known Y dwarfs at 4.5 μm and two to five magnitudes (a factor of 6 to 100) redder in $J - [4.5]$ color. Model analysis shows that it is very cold, with $T_{eff} \approx 250$ K (Figure 1; Beamin et al. 2014; Leggett et al. 2015, 2017; Luhman 2014; Schneider et al. 2016; Zapatero Osorio et al. 2016). Evolutionary models indicate an age of about 3 Gyr and a mass of only about five times that of Jupiter (Leggett et al. 2017). This cold and small object is very faint and has a luminosity 100 million times fainter than the Sun and only 10 times greater than Jupiter's.

Figure 5 shows the measured brightness of WISE 0855 as a function of wavelength, using imaging data taken with filters covering the far-red to the mid-infrared. The best model fit as of April 2017 is also shown. The model is successful where most of the flux is emitted, in the mid-infrared and also at *Y* and *J*. However all current models (at the time of writing) produce too much flux at wavelengths around 0.9 μm and too little flux at *H* and [3.6], to various degrees. Clearly this unique object remains a challenge, and it is a high priority target for the soon-to-fly *James Webb Space Telescope* (Greenhouse 2016).

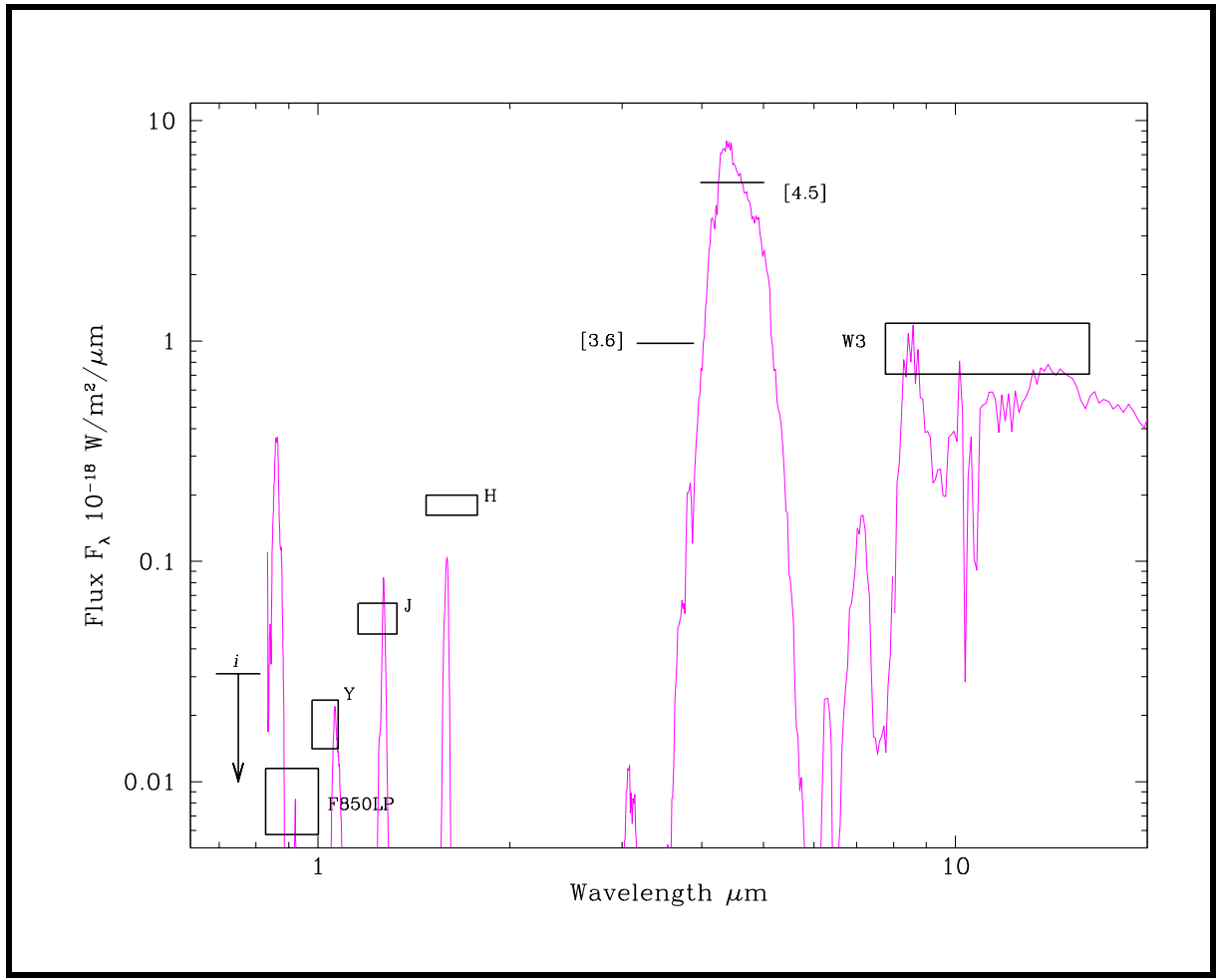

**Fig. 5** Black lines and boxes indicate the brightness of WISE J085510.83−071442.5 (Luhman *2014*) in various filters as indicated by the legends, scaled for a distance of 10 pc. The model shown is the best fit as of April 2017 and is a solar-metallicity cloud-free model from Tremblin et al. (*2015*) which includes vertical mixing and nonequilibrium chemistry. The model has $T_{\rm eff}$ = 250 K and a surface gravity $g = 10^{3.5}$ cm s$^{-2}$. Note that the axes are logarithmic. Based on Figure 19 of Leggett et al. (*2017*).

# How Can We Find More Cold Brown Dwarfs?

Brown dwarfs with a range in mass will have formed throughout the lifetime of our Galaxy, in molecular clouds with a range of chemical composition. A 5-Jupiter-mass brown dwarf that was formed 10 Gyr ago will have cooled to 170 K. The current sample of Y dwarfs is composed of 16 objects with 375 < $T_{\rm eff}$ K < 475 and 8 with 250 < $T_{\rm eff}$ K < 350 (Leggett et al. *2017*). This is a statistically small sample, which is unlikely to represent the diversity of the actual Y dwarf population. All the known Y dwarfs with $T_{\rm eff} \approx 300$ K Y are within 10 pc. Assuming a uniform space density, a doubling of the distance to 20 pc will increase this number by a factor of 8 and give a good sample of about 60 cold objects. This assumes full sky coverage of 41,000 square degrees, as achieved by *WISE*.

A 300 K Y dwarf with an age similar to our Sun (5 Gyr) at a distance of 20 pc will have an *H* (1.6 μm) magnitude of 26.5 and a 4.5 μm (IRAC C2 or *WISE* W2) magnitude of 17.5 (or equivalently 3e-8 Jy at 1.6 μm and 2e-5 Jy at 4.5 μm). This is much fainter than reached by

*WISE* or upcoming wide-field space missions such as EUCLID (Racca et al. *2016*) or WFIRST (Gehrels et al. *2015*). Although *Spitzer* and the *James Webb Space Telescope* can reach these sensitivities, neither is intended to cover large areas of sky; the *Spitzer* Legacy and Exploration Science wide-field surveys cover a few tens of square degrees of sky, for example.

The 5 μm wavelength region is the most practical for detecting cold brown dwarfs (Figures 1, 2, 5; Burrows et al. *2003*). In order to find more of these cold Jupiter-size objects, a new space-based infrared sky survey is needed. The telescope would be similar in size to *Spitzer*, with a mirror diameter ≈0.9 m (twice the size of the *WISE* mirror). The entire sky would need to be mapped in at least two filters, centered near 3.5 and 4.5 μm, i.e., similar to the IRAC C1 and C2 or the *WISE* W1 and W2 filters. These two filters allow the identification of low temperature objects which have strong $CH_4$ absorption at 3 μm (Figs. 1, 2, and 5). The time frame for the mission would be similar to the time needed for *WISE* to scan the entire sky twice, 1–2 years. Of course, a mission with a larger mirror that can detect more distant cold brown dwarfs could cover a smaller sky area, in a shorter amount of time. However, having images taken separated by a year or more also allows the identification of nearby fast-moving objects, as was the case for WISE 0855 (Luhman *2014*).

# -Update, 2023

## Population

In 2017, 24 Y dwarfs were known, 16 with 375 < $T_{eff}$ K < 475 and 8 with 250 < $T_{eff}$ K < 350. The population has doubled in 2023, thanks to the NEOWISE survey which was reactivated in 2013 and continues in 2023 (Mainzer et al. 2014). NEOWISE has increased the depth of the all-sky WISE survey at wavelength of 3.4 and 4.6 μm. The multiple imaging epochs also enables searches for faint and red sources with significant proper motion (Eisenhardt et al. 2020, Meisner et al. 2020).

As of September 2023, 50 Y dwarf systems are known (Leggett et al. 2021, Table 6; see also Kirkpatrick et al. 2021; and references therein). One of these, WISE J033605.05–014350.4, has been resolved into the first confirmed Y dwarf binary system, using the James Webb Space Telescope (JWST; Calissendorff et al. 2023). Of the 51 sources, 12 are cooler than 375K. The coldest remains WISE J085510.83−071442.5 with $T_{eff}$ = 260K (Luhman 2014, Leggett et al. 2021), followed by the secondary of the WISE J033605.05–014350.4 system, which has $T_{eff} \approx$ 300K (Calissendorff et al. 2023, Leggett & Tremblin 2023). Assuming that these two Y dwarfs have ages typical of the solar neighborhood, these temperatures imply very low masses of 5 to 10 Jupiter-masses.

The extension of the NEOWISE mission also revealed a population of high proper motion dwarfs with colors indicating that they are extremely metal poor. The metal paucity and large motion suggests that these are old brown dwarfs. One member of this population may be cold enough to be a Y dwarf, but that remains to be confirmed (Meisner et al. 2023).

## Model Atmospheres and Synthetic Spectra

Advances have been made in modelling Y dwarf atmospheres. In 2017 it was suggested that the systematic discrepancies between observed and modelled colors and spectra for cooler Y dwarfs may be addressed by including $H_2O$ clouds. However recent atmospheric circulation models of brown dwarfs and exoplanets which include rapid rotation show that rotation has a large impact on the energy transport through the atmosphere (Tan & Showman 2021). Y dwarfs are rapid rotators (Hsu et al. 2021). Leggett et al. (2021) generated models with empirical adjustments to the pressure:temperature ($P:T$) relationship, which maps the cooling of the atmosphere from the core to the surface. Those empirical but physically motivated models, known as ATMO2020++, produce synthetic spectra which significantly and comprehensively improve the agreement with observations, for late-T and Y dwarfs. The first JWST spectrum of a Y dwarf (Beiler et al. 2023) validates the ATMO2020++ models for $T_{eff}$ = 450K with the remaining systematic discrepancies occuring at $T_{eff}$ < 350K (Leggett & Tremblin 2023). At those very cold temperatures we may indeed be seeing the impact of $H_2O$ clouds in the upper atmosphere (Morley et al. 2014, Lacy & Burrows 2023).

## The Next Mid-Infrared Space Mission

In December 2022 NASA approved the NEO Surveyor Project for flight, with launch expected in 2028 (https://neos.arizona.edu/). The Project is optimized for the detection of potentially hazardous asteroids, and will obtain multiple-epoch images of about one-quarter of the sky at wavelengths of 4.5 μm and 8 μm (Mainzer et al. 2015). Y dwarfs are bright at 4.5 μm and faint at 8 μm, so they can be identified in the Project data by their unusual color, and also by their motion. The nominal survey region and sensitivity (Mainzer et al. 2015, Sonnett et al. 2021) suggests that only a small number of new Y dwarfs will be identified (Leggett et al. 2019). However, if the NEO Surveyor is as successful and long-lived as the WISE and NEOWISE missions, it may surpass these expectations and reveal new populations of these fascinating, cold, very low mass, solar neighbors.

# Conclusion (Updated 2023)

Space-based missions which have imaged the sky at $\lambda \approx$ 5 μm have discovered a population of cold, planet-like, brown dwarfs near our Sun. Discovery and characterization of these so-called Y-type brown dwarfs have made significant progress since their discovery in 2011. Expanding the known population to the current 51 objects (as of September 2023) required extending the duration of the *Wide-field Infrared Survey Explorer* (*WISE)* mission and the development of techniques to extract faint, red, and moving sources from the large database. Characterization required more complex atmospheric models which include condensation, turbulent mixing of gas, and changes brought about by rapid rotation. Moreover, these models had to span the range in surface gravity, chemical composition, and temperature, seen in the expanding Y dwarf population.

Although great progress has been made, only 12 Y dwarfs are known which are cooler than 375K, and model atmospheres show systematic deviations from the observed colors for the coldest objects, possibly due to the onset of water clouds. It is important to expand the sample of isolated, very cold and nearby bodies to better understand the environment of our Sun, and the processes of star and planet formation. A new mid-infrared space project, called the NEO Surveyor, is expected to launch in 2028. Optimized for the detection of hazardous asteroids, the camera filters are nevertheless suitable for the discovery of Y dwarfs. The nominal size and sensitivity of the survey suggests that not many new Y dwarfs will be found, however if the NEO Surveyor is as successful and long-lived as the WISE and NEOWISE missions, it may surpass these expectations.

# References

Ackerman AS, Marley MS (2001) Precipitating condensation clouds in substellar atmospheres. ApJ 556:872

Baraffe I, Chabrier G, Allard F, Hauschildt PH (2002) Evolutionary models for low-mass stars and brown dwarfs: uncertainties and limits at very young ages. A&A 382:563

Beamin JC, Ivanov VD, Bayo A et al (2014) Temperature constraints on the coldest brown dwarf known: WISE 0855-0714. A&A 570:L8

Beiler SA, Cushing MC, Kirkpatrick JD et al (2023) The First JWST Spectral Energy Distribution of a Y Dwarf. ApJ 951:L48

Bowler BP, Hillenbrand LA (2015) Near-infrared spectroscopy of 2M0441+2301 AabBab: a quadruple system spanning the stellar to planetary mass regimes. ApJ 811:L30

Burgasser AJ, Kirkpatrick JD, Brown ME et al (1999) Discovery of four field methane (T-type) dwarfs with the two micron all-sky survey. ApJ 522:L65

Burrows A, Marley MS, Hubbard WB et al (1997) A nongray theory of extrasolar giant planets and brown dwarfs. ApJ 491:856

Burrows A, Hubbard WB, Lunine JI, Liebert J (2001) The theory of brown dwarfs and extrasolar giant planets. RvMP 73:719

Burrows A, Sudarsky D, Lunine JI (2003) Beyond the T dwarfs: theoretical spectra, colors, and detectability of the coolest brown dwarfs. ApJ 596:587

Calissendorff P, De Furio M, Meyer M et al (2023) JWST/NIRCam Discovery of the First Y+Y Brown Dwarf Binary: WISE J033605.05–014350.4. ApJL 947:L30

Cannon AJ, Pickering EC (1912) Classification of 1,477 stars by means of their photographic spectra. AnHar 56:65

Chabrier G (2003) Galactic stellar and substellar initial mass function. PASP 115:763

Cushing MC, Kirkpatrick JD, Gelino CR et al (2011) The discovery of Y dwarfs using data from the wide-field infrared survey explorer (WISE). ApJ 743:50

Cushing MC, Hardegree-Ullman KK, Trucks JL et al (2016) The first detection of photometric variability in a Y dwarf: WISE J140518.39+553421.3. ApJ 823:152

Eisenhardt PRM, Marocco F, Fowler, JW et al (2020) The CatWISE Preliminary Catalog: Motions from WISE and NEOWISE Data. ApJS 247:69.

Fazio GG, Hora JL, Allen LE et al (2004) The infrared array camera (IRAC) for the spitzer space telescope. ApJS 154:10

Gehrels N, Spergel D, WFIRST SDT Project (2015) Wide-field infraRed survey telescope (WFIRST) mission and synergies with LISA and LIGO-virgo. J Phys Conf Ser 610:2007

Greenhouse MA (2016) The JWST science instrument payload: mission context and status. SPIE 9904:6

Hayashi C, Nakano T (1963) Evolution of stars of small masses in the pre-main-sequence stages. Prog Theor Phys 30:460

Hsu C-C, Burgasser AJ, Theissen CA et al (2021) The Brown Dwarf Kinematics Project (BDKP). V. Radial and Rotational Velocities of T Dwarfs from Keck/NIRSPEC High-resolution Spectroscopy. ApJS 257:45.

Kirkpatrick JD, Reid IN, Liebert J et al (1999) Dwarfs cooler than "M": the definition of spectral type "L" using discoveries from the 2 micron all-sky survey (2MASS). ApJ 519:802

Kirkpatrick JD, Gelino CR, Cushing MC et al (2012) Further defining spectral type "Y" and exploring the low-mass end of the field brown dwarf mass function. ApJ 753:156

Kirkpatrick JD, Gelino CR, Faherty JK et al (2021) The Field Substellar Mass Function Based on the Full-sky 20 pc Census of 525 L, T, and Y Dwarfs. ApJS 253:7.

Kroupa P (2001) On the variation of the initial mass function. MNRAS 322:231

Kumar SS (1967) On planets and black dwarfs. Icarus 6:136

Lacy B & Burrows A (2023) Self-consistent Models of Y Dwarf Atmospheres with Water Clouds and Disequilibrium Chemistry. ApJ 950:8

Leggett SK, Golimowski DA, Fan X et al (2002) Infrared photometry of late-M, L, and T dwarfs. ApJ 564:452

Leggett SK, Morley CV, Marley et al (2013) A comparison of near-infrared photometry and spectra for Y dwarfs with a new generation of cool cloudy models. ApJ 763:130

Leggett SK, Morley CV, Marley MS, Saumon D (2015) Near-infrared photometry of Y dwarfs: low ammonia abundance and the onset of water clouds. ApJ 799:37

Leggett SK, Cushing MC, Hardegree-Ullman KK et al (2016) Observed variability at 1 and 4 μm in the Y0 brown dwarf WISEP J173835.52+273258.9. ApJ 830:141

Leggett SK, Tremblin P, Esplin TL, Luhman KL, Morley CV (2017) The Y-type brown dwarfs: estimates of mass and age from new astrometry, homogenized photometry, and near-infrared spectroscopy. ApJ 842:118

Leggett S, Apai D, Burgasser A et al (2019) Discovery of Cold Brown Dwarfs or Free-Floating Giant Planets Close to the Sun. Astro2020: Decadal Survey on Astronomy and Astrophysics, science white papers, no. 95; Bulletin of the American Astronomical Society, Vol. 51, Issue 3, id. 95

Leggett SK, Tremblin P,, Phillips MW et al (2021) Measuring and Replicating the 1–20 $\mu$m Energy Distributions of the Coldest Brown Dwarfs: Rotating, Turbulent, and Nonadiabatic Atmospheres. ApJ 918:11.

Leggett SK & Tremblin P (2023) The First Y Dwarf Data From JWST Show That Dynamic and Diabatic Processes Regulate Cold Brown Dwarf Atmospheres. ApJ submitted.

Li Y, Kouwenhoven MBN, Stamatellos D, Goodwin SP (2015) The dynamical evolution of low-mass hydrogen-burning stars, brown dwarfs, and planetary-mass objects formed through disk fragmentation. ApJ 805:116

Lord SD (1992) A new software tool for computing Earth's atmospheric transmission of near- and far-infrared radiation. NASA Tech Memo 103957

Luhman KL (2014) Discovery of a ∼250 K brown dwarf at 2 pc from the sun. ApJ 786:L18

Mainzer A, Cushing MC, Skrutskie M et al (2011) The first ultra-cool brown dwarf discovered by the wide-field infrared survey explorer. ApJ 726:30

Mainzer A, Bauer J, Cutri RM et al (2014) Initial Performance of the NEOWISE Reactivation Mission. ApJ 792:30

Mainzer A, Grav T, Bauer J, et al. (2015) Survey Simulations of a New Near-Earth Asteroid Detection System. AJ 149:172

Meisner, AM, Caselden D, Kirkpatrick JD et al (2020) Expanding the Y Dwarf Census with Spitzer Follow-up of the Coldest CatWISE Solar Neighborhood Discoveries. ApJ 889:74.

Meisner AM, Leggett SK, Logsdon SE et al (2023) Exploring the Extremes: Characterizing a New Population of Old and Cold Brown Dwarfs. AJ 166:57

Miller GE, Scalo JM (1979) The initial mass function and stellar birthrate in the solar neighborhood. ApJS 41:513

Morgan WW, Keenan PC (1973) Spectral classification. ARA&A 11:29

Morley CV, Fortney JJ, Marley MS et al (2012) Neglected clouds in T and Y dwarf atmospheres. ApJ 756:172

Morley CV, Marley MS, Fortney JJ et al (2014) Water clouds in Y dwarfs and exoplanets. ApJ 787:78

Racca GD, Laureijs R, Stagnaro L et al (2016) The Euclid mission design. SPIE 9904:23 pp. https://doi.org/10.1117/12.2230762

Salpeter EE (1955) The luminosity function and stellar evolution. ApJ 121:161

Saumon D, Geballe TR, Leggett SK et al (2000) Molecular abundances in the atmosphere of the T dwarf GL 229B. ApJ 541:374

Saumon D, Marley MS (2008) The evolution of L and T dwarfs in color-magnitude diagrams. ApJ 689:1327

Saumon D, Marley MS, Abel M, Frommhold L, Freedman RS (2012) New $H_2$ collision-induced absorption and $NH_3$ opacity and the spectra of the coolest brown dwarfs. ApJ 750:74

Schneider AC, Cushing MC, Kirkpatrick DJ, Gelino CR (2016) The collapse of the wien tail in the coldest brown dwarf? Hubble space telescope near-infrared photometry of WISE J085510.83-071442.5. ApJ 823:35

Showman A, Kaspi Y (2013) Atmospheric dynamics of brown dwarfs and directly imaged giant planets. ApJ 776:85

Sonnett S, Mainzer A, Grav T, et al (2021) NEO Surveyor Cadence and Simulations. 7th IAA Planetary Defense Conference, held in Vienna, Austria, 26-30 April 2021, id. 56

Spiegel DS, Burrows A (2012) Spectral and photometric diagnostics of giant planet formation scenarios. ApJ 745:174

Strauss MA, Fan X, Gunn JE et al (1999) The discovery of a field methane dwarf from sloan digital sky survey commissioning data. ApJ 522:L61

Tan X & Showman AP (2021) Atmospheric circulation of brown dwarfs and directly imaged exoplanets driven by cloud radiative feedback: effects of rotation. MNRAS 502:678.

Tremblin P, Amundsen DS, Mourier P et al (2015) Fingering convection and cloudless models for cool brown dwarf atmospheres. ApJ 804:L17 (T15)

Tsuji T, Ohnaka K, Aoki W, Nakajima T (1996) Evolution of dusty photospheres through red to brown dwarfs: how dust forms in very low mass objects. A&A 308:L29

Warren SJ, Mortlock DJ, Leggett SK et al (2007) A very cool brown dwarf in UKIDSS DR1. MNRAS 381:1400

Wright EL, Eisenhardt PRM, Mainzer AK et al (2010) The wide-field infrared survey explorer (WISE): mission description and initial on-orbit performance. AJ 140:1868

Yurchenko SN, Barber RJ, Tennyson J (2011) A variationally computed line list for hot $NH_3$. MNRAS 413:1828

Yurchenko SN, Tennyson J (2014) ExoMol line lists – IV. The rotation-vibration spectrum of methane up to 1500 K. MNRAS 440:1649

Zahnle KJ, Marley MS (2014) Methane, carbon monoxide, and ammonia in brown dwarfs and self-luminous giant planets. ApJ 797:41

Zapatero Osorio MR, Lodieu N, Bejar VJS et al (2016) Near-infrared photometry of WISE J085510.74-071442.5. A&A 592:80

Zinnecker H, Yorke HW (2007) Toward understanding massive star formation. A&A 45:481